\documentclass[twocolumn,prc,showkeys,showpacs]{revtex4}
\usepackage{graphicx} % Include figure files
\usepackage{amsmath}

\newcommand{\up}[2]{$^{#1#2}$}
\newcommand{\updown}[4]{$_{#3#4}^{#1#2}$}
\newcommand{\Qspec}{Q_{\rm spec}}
\newcommand{\BibTitle}[1]{#1}
\renewcommand{\BibTitle}[1]{}
\newcommand{\mbibitem}[1]{\bibitem#1}
\renewcommand{\mbibitem}[1]{\bibitem}
\newcommand{\parp}{$^+$}
\newcommand{\parm}{$^-$}

\begin{document}
\title{The Cranked Nilsson-Strutinsky versus the Spherical Shell
Model:\\
A Comparative Study of pf-Shell Nuclei}
\author{Andrius Juodagalvis$^{1,2)}$}
%\email{andrius@theory.gsi.de}
\author{Ingemar Ragnarsson$^{1)}$}
%\email{ingemar.ragnarsson@matfys.lth.se}
\author{Sven {\AA}berg$^{1)}$}
%\email{sven.aberg@matfys.lth.se}
\affiliation{$^{1)}$Mathematical Physics, Lund Institute of Technology \\
P.O.\ Box 118, S-221 00 Lund, Sweden}
\affiliation{$^{2)}$GSI,
Planckstr.\ 1, 64291 Darmstadt, Germany}
\date{\today}

\begin{abstract}
A comparative study is performed of a deformed mean field theory,
represented by the cranked Nilsson-Strutinsky (CNS) model, and the
spherical shell model. Energy spectra, occupation numbers,
$B(E2)$-values, and spectroscopic quadrupole moments in the light $pf$
shell nuclei are calculated in the two models and compared. The result
is also compared to available experimental data which are generally
well described by the shell model. Although the Nilsson-Strutinsky
calculation does not include pairing, both the subshell occupation
numbers and quadrupole properties are found to be rather similar in
the two models. It is also shown that ``unpaired'' shell model
calculations produce very similar energies as the CNS at all
spins. The role of the pairing energy in the description of
backbending and signature splitting in odd-mass nuclei is also
discussed.
\end{abstract}



\maketitle

\section{Introduction}
A large number of models have been developed to get insight into
the spectroscopic properties of nuclei. Two of the most successful
models are the spherical shell model and the deformed shell model.
Large-scale spherical shell model calculations provide excellent
agreement with observed data, but less transparent physics
interpretations. The deformed shell model, on the other hand, is a
mean field approach that is more illustrative but gives a less
accurate agreement with data. A parallel study using the two
models allows a better understanding of the underlying nuclear
processes. In this paper we aim at comparing the predictions of
the spherical shell model (SM) and one version of the deformed
shell model, namely, the configuration-dependent cranked
Nilsson-Strutinsky (CNS) approach.

For nuclei in the mass region $A\sim40$-50 with valence particles
occupying $pf$-shell orbits, well-tested shell model calculations are
available \cite{Caurier04}. Nuclei in this region show several
interesting collective phenomena, such as the existence of rotational
bands, backbending of the yrast band, band termination, and the
appearance of superdeformed as well as axially asymmetric shapes
(triaxiality) \cite{Brandolini04,Ur04,Lenzi02}. Other interesting
features which have been discussed for these nuclei are the role of
isoscalar and isovector pairing \cite{Caurier04,Satula01b}, violation
of SU(3) symmetry \cite{Gueorguiev00,Poves04}, violation of isospin
symmetry \cite{Zuker02,Garrett01}, angular momentum dependence of the
mirror energy difference
\cite{OLeary02,Zuker02,Tonev02,Lenzi01,Bentley98,Bentley99}, and
Jacobi shapes \cite{Maj04}. This long, by no means complete list
indicates the significant attention that these nuclei received in the
last few years.

Though both the spherical shell model and the cranked
Nilsson-Strutinsky model provide a microscopic description of a
nucleus, they are basically different. One difference is that the
shell model gives a laboratory frame description, while the CNS
provides a description in the intrinsic frame of reference.  Another
difference is the 'model space' and the treatment of the nuclear
interaction.  Within the restricted model space of the spherical shell
model, the residual interaction between the valence particles is
completely taken into account.  The deformed shell model uses a
virtually unrestricted model space.  However, only specific parts of
the nuclear interaction are included, in particular the
quadrupole-quadrupole interaction. Furthermore, the inclusion of this
interaction is made in the mean-field approximation, with the
self-consistency condition treated in an approximate way through the
Strutinsky energy theorem \cite{Strutinsky68}.  Comparison of the two
models allows to identify the missing parts of the nuclear interaction
as well as correlations beyond the mean field in the deformed shell
model on one hand, and the model space limitations in the spherical
shell model on the other hand.

The development of the shell model computer code {\sc Antoine}
\cite{Antoine} led to extensive and rather systematic theoretical
investigations of nuclear spectroscopy in the lower part of $pf$ shell
\cite{A48SM,Cr50SM,A47A49SM,Mn50SM} and inspired much experimental
work. The (unpaired) cranked Nilsson-Strutinsky model
\cite{CNSpar,Afanasjev99} has been applied to more or less all nuclei
in the periodic table for which high-spin states have been
studied. The model successfully describes terminating rotational bands
\cite{Afanasjev99}, superdeformed bands \cite{Rag93,Afa96,Afa99} and
the phenomenon of shape coexistence \cite{Ma02}. This model has also
been used to describe a few nuclei in the region of the present
study. The interpretation of selected high-spin data in
$^{47,48,49}$Cr and $^{47}$V was discussed in Refs.\
\cite{Cr48triax,Afanasjev99}, while the odd-odd nuclei \up46V and
\up50Mn were investigated in Ref.\ \cite{Dong03}.

The spherical shell model was previously compared to the CNS model for
some nuclei in the $A \approx 40$ region \cite{Cr48triax,Ar36,Dong03},
and a striking similarity in the predictions was found. In this paper
we enlarge the scope of comparison between the two types of models to
more nuclei and discuss even-even as well as odd-$A$
nuclei. Backbending, the role of pairing, and in particular its
contribution to the signature splitting are stressed. We shall
concentrate on the $pf$-shell nuclei close to the $N=Z$ line with mass
numbers between $A=44$ and 49. By making a more systematic comparison
of the two models, we want to improve our understanding of the
deformed mean field model, test its validity for fairly light nuclei,
and, in particular, to get a better understanding of the physical
picture behind the observed properties in the region. This allows to
assess the reliability of the CNS approach in heavier nuclei, where
spherical shell model calculations are not feasible at present.

The paper is organized in the following way. First, we describe
the models in section \ref{sect-Models}. Predictions of the two models
are confronted and compared in section \ref{sect-Results}. The paper
is summarized in section \ref{sect-Summary}.

% -------------------------------------------
\section{Models}
\label{sect-Models}
\subsection{The shell model and the cranked Nilsson-Strutinsky model}
\label{subsect-Models}

Shell model results are calculated using the computer code {\sc
Antoine} \cite{Antoine}. Valence particles, occupying orbits in
the full $pf$ shell, interact via the residual interaction KB3
\cite{KB3}. For quadrupole properties, effective charges
$e_p=1.5e$, $e_n=0.5e$ are used. Most of the results obtained
with a ``full'' interaction have already appeared in the
publications by the Madrid-Strasbourg group (for a recent review
see Ref.\ \cite{Caurier04}). Here we will also present "unpaired
calculations" which, with exception of the nucleus \up48Cr
\cite{Cr48pair}, have not been presented before.

CNS calculations are performed utilizing the modified oscillator
potential and a standard set of parameters
\cite{Afanasjev99,CNSpar}. The implementation allows to minimize
the energy for a fixed configuration at a given value of the total
angular momentum with configurations defined as explained in
Refs.\ \cite{CNSconf,Afanasjev99}. The total energy is minimized
varying three degrees of freedom: two quadrupole parameters,
$\varepsilon$ (deformation) and $\gamma$ (non-axiality), and one
hexadecapole parameter, $\varepsilon_4$ \cite{CNSpar}. All kinds
of pairing interactions are neglected in the CNS approach.
Therefore, the CNS results are mainly valid at high spin, and it
becomes natural to normalize experimental and calculated energies
at some high spin value. This is contrary to the spherical shell
model where the corresponding normalization is generally done at
the ground state.

The calculated energies are often plotted with a subtracted
rotational reference $E_{\rm ref}=32.32A^{-5/3}I(I+1)$ MeV
\cite{Afanasjev99} in order to facilitate reading of figures and
to highlight differences relative to this rotational behavior. The
reference corresponds to the rotation-energy of a rigid rotor for
a prolate nucleus with a radius constant $r_0=1.20$ fm and a
deformation $\varepsilon\approx0.23$.

\subsection{Moments}
\label{subsect-Moments} A translation between the intrinsic frame
of reference and the laboratory frame of reference can be obtained
using the rotor model, where those two frames can be related, see
\cite{BM2}. Thus, having values of the quadrupole deformation
parameters, $\varepsilon$ and $\gamma$, it is possible to estimate
the strength of the $E2$ transition between two states, $B(E2)$,
and the spectroscopic quadrupole moment of a state, $\Qspec$. And
vice versa, using the values of $B(E2)$ and $\Qspec$, the
quadrupole deformation of a nucleus can be derived. Formally, this
identification is valid only for fixed axially symmetric shapes.

In the CNS approach, we calculate the intrinsic quadrupole moments
from proton single-particle wave functions at appropriate equilibrium
deformations. Neutrons have no contribution to this moment. Since the
rotor model assumes axially symmetric shape, while the calculated
triaxiality parameter is usually sizeable, we use an approximate
relation between the intrinsic moments and laboratory-frame
observables \cite{Cr48triax}.  If a nucleus has a rather flat energy
surface, quantum fluctuations may become important. To show their
effect, in a few selected cases we utilize an approximate method
\cite{Cr48triax} to add the effect of quantum fluctuations on
calculated quadrupole properties.

\subsection{Occupation numbers}
\label{subsect-Occ} Single $j$-shell occupation numbers are
readily obtained from the shell model wave-function. In the CNS
calculations, the eigenstates are expanded in a stretched basis.
It is however
straightforward to make a transformation into a spherical basis
and then to add up the fractions of the
spherical subshells of the eigenstates at the equilibrium deformations.
The method is outlined in Ref.\ \cite{CNSoccnum}.

\subsection{Pairing}
\label{subsect-Pairing}
The shell model includes all kinds of correlations, an important part
of which is pairing. To investigate the effect of the pairing
interaction, two calculations are performed: one using the full
interaction, and another one using the interaction with the pairing
force subtracted. As defined in Ref.\ \cite{Cr48pair}, the pairing
energy is the difference between the energies obtained in those two
calculations. We consider both the isoscalar ($np$; $J=1,$ $T=0$) and
the isovector ($nn$+$pp$+$np$; $J=0,$ $T=1$) $L=0$ pairing
\cite{Cr48pair,DufourZuker}. Thus three kinds of pairing energy are
discussed: $T=0$ pairing energy, which is deduced from the interaction
with a subtracted $T=0$ pairing force; $T=1$ pairing energy, which is
deduced from the interaction with a subtracted $T=1$ pairing force;
and the full pairing energy, which is deduced from the interaction
with both pairing forces subtracted. The strengths of the forces are
\cite{DufourZuker}: $G_{T=0}=-0.51\hbar\omega$ and $G_{T=1}
=-0.32\hbar\omega$, where $\hbar\omega=40A^{-1/3}$ MeV. Since the CNS
approach does not include the pairing interaction, it is reasonable to
compare its predictions to the ``unpaired'' energies of the shell
model.

% -----------------------------------------------------------------
% -----------------------------------------------------------------
\section{Results}
\label{sect-Results} We compare the two models for the even-even
systems in subsection \ref{ssect-EENuclei}, where the nuclei
${}_{\phantom{44,}22}^{44,46}$Ti and \updown4824Cr are discussed.
For \up48Cr, the negative-parity band is discussed in addition to
the ground-state band. A few selected odd-even nuclei (namely,
\updown4522Ti, \updown4723V and \updown4924Cr) are discussed in
subsection \ref{ssect-OENuclei}.

% ---------------------------
\subsection{Even-even nuclei}
\label{ssect-EENuclei}

% ---------------------------
\subsubsection{Positive-parity band in \up48Cr}
\label{sssect-Cr48}
The nucleus \up48Cr has a half-filled $f_{7/2}$
shell of protons and neutrons, resulting in the largest ground-state
deformation in the $f_{7/2}$ region. The yrast band shows an
interesting behavior, being rotor-like with a backbend and a
well-established termination at $I=16^+$. It has been interpreted as
having a triaxial shape \cite{Cr48comp,Cr48triax}. The pairing energy along the
band as well as quadrupole properties have been studied extensively
\cite{Cr48pair,Cr48comp,Gueorguiev00,Cr48Exp,A48SM,Cr48BT,Cr48Hara,Cr48Tanaka}.
It is, therefore, a good example to start this broader comparison
between the CNS and the shell model.

\begin{figure}[tbp]
\includegraphics*[width=7.5cm, angle=0]{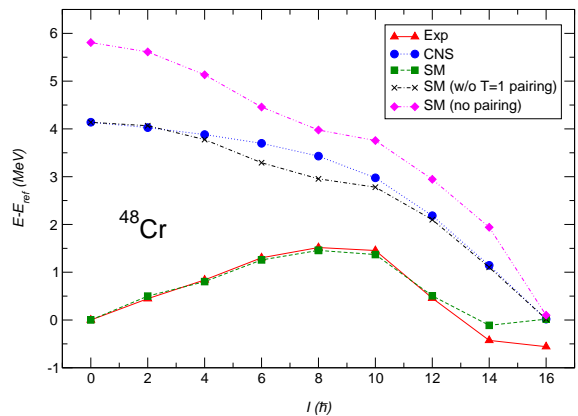}
\caption{ (Color online)
Energies in the \up48Cr yrast band plotted relative
to a rotational reference. Experimental values \cite{Cr48Exp}
are shown by triangles. The other four lines show calculated
results: CNS (circles), full shell model (squares), and two
``unpaired'' shell model cases: without the $T=1$ pairing (crosses) and
without both $T=0$ and $T=1$ pairings (diamonds).
The calculated CNS energies
are matched to the excitation energy of the fully aligned 16\parp\
state.
}
\label{fig-Cr48ERef}
\end{figure}

An exploratory study of \up48Cr using the CNS and the shell model
approaches \cite{Cr48triax} suggested that the predicted
quadrupole properties are similar, although energies are rather
different, see Fig.\ \ref{fig-Cr48ERef}. (A similar conclusion was
reached when using the cranked Hartree-Fock-Bogolyubov method
\cite{Poves04}.) We continue this comparison particularly
emphasizing the role of the pairing interaction. As in previous
studies, we concentrate on the yrast positive-parity, even-spin
states between $I=0$ and 16.

%----------
%----------

Since pairing is neglected in the CNS calculation, Fig.\
\ref{fig-Cr48ERef} also shows energies of two ``unpaired'' shell model
calculations. The change in the excitation energies suggests that the
backbending behavior is mainly caused by the $T=1$ pairing (as pointed
out in \cite{Cr48pair} and further discussed in \cite{Cr48BT}), while
the $T=0$ pairing is generally smaller and decreases smoothly with
spin. When the CNS energies are normalized to match the excitation
energy of a fully aligned state at $I=16^+$, they fall between the two
unpaired shell model energies; with only the $T=1$ pairing neglected
and all pairing neglected completely, respectively. However, they
come much closer to the shell model results without the $T=1$ pairing.
A similar behavior was noted in Ref.\ \cite{Cr48comp}, where the shell
model and CHFB calculations were compared. There it was attributed to
the improper treatment of the proton-neutron pairing by the latter
mean-field method \cite{Cr48comp,Poves04}.

\begin{figure}[tbp]
\includegraphics*[width=7.5cm, angle=0]{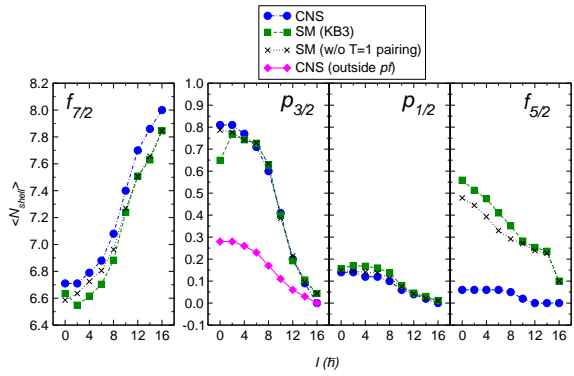}
\caption{ (Color online) Average number of particles (neutrons and
protons) in the {\em spherical\/} $j$-shells in \up48Cr: the shell
model values calculated using the full interaction and the
interaction without the $T=1$ pairing are shown by squares and
crosses, respectively. The CNS occupations of the $pf$ shell model
orbitals are shown by circles, while the occupations of orbitals
outside the shell model space are shown by diamonds in the panel
with the $p_{3/2}$ shell occupancies. } \label{fig-Cr48Occ}
\end{figure}

We already noted the similarity of the quadrupole properties as
described by the two approaches \cite{Cr48triax} and will discuss
them more below. Here we would like to point out that a more
detailed investigation of the results shows that the calculated
wave functions are similar as well: The spherical $j$-shells are
occupied almost identically, see Fig.\ \ref{fig-Cr48Occ}. In
particular, there is a very good agreement in the occupation of
the $f_{7/2}$, $p_{3/2}$ and $p_{1/2}$ shells, despite the fact
that the two models treat the ``model space'' in a different
manner, and moreover, pairing is completely neglected in the CNS.
The effects of the $T=1$ pairing interaction are visible in the
increased occupation of the $f_{5/2}$ shell and the decreased
occupation of the $f_{7/2}$ shell. This is easily understood in
the BCS picture of the $T=1$ pairing. The pairing causes
occupations of the orbitals around the Fermi surface to be smeared
out, and in \up48Cr the Fermi surface is in the middle of the $f_{7/2}$
shell and below the other shells. 

The CNS predicts a much lower occupation of the $f_{5/2}$ shell than
the shell model (0.06 particles in the ground state as compared to
0.56 particles predicted by the shell model). In addition, some
occupation is found outside of the $pf$ shell.  The contents of
excitations beyond the spherical shell model space decreases
smoothly with spin from 0.28 particles in the ground state to zero in
the band-terminating state.  This agreement in occupation numbers
between the two models is remarkable, since partial occupancies of the
spherical $j$-shells have different origin in the two models.  The
two-body interaction between valence particles causes configuration
mixing in the shell model. On the other hand, the mixing of spherical
$j$-shells is determined by the deformation and rotation in the CNS
approach.

\begin{figure}[tbp]
\includegraphics*[width=7.5cm, angle=0]{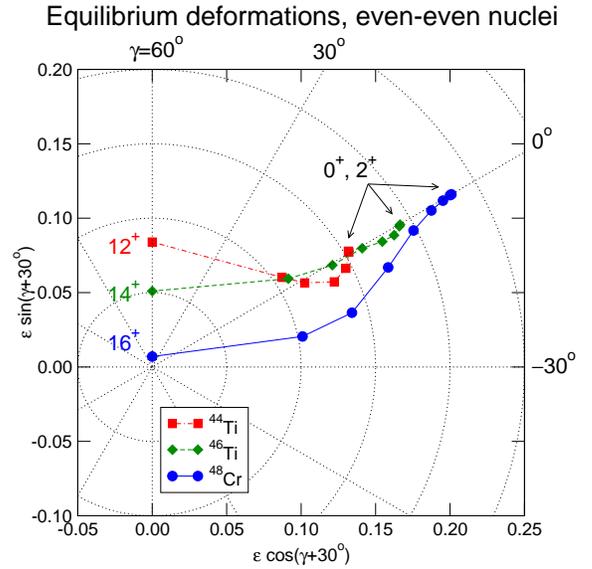}
\caption{ (Color online)
CNS calculated equilibrium deformations along the yrast
bands in three even-even nuclei: \up44Ti (squares), \up46Ti (diamonds)
and \up48Cr (circles).
The deformation change between the $I=0$ and $I=2$ states is
negligible in all these nuclei.
}
\label{fig-EE-EqConf}
\end{figure}

\begin{figure}[tbp]
\includegraphics*[width=7.5cm, angle=0]{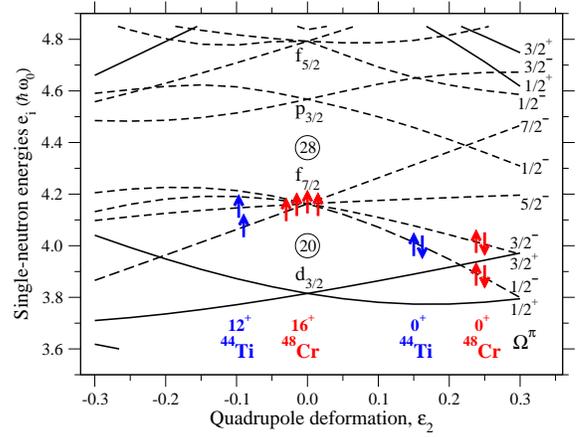}
\caption{ (Color online)
A Nilsson diagram for single-neutron states in the modified
oscillator potential. The arrows symbolize particles,
while their direction indicates whether $m_j$ is positive (arrow
up) or negative (arrow down). The ground state as well as the
band-terminating state with all particles in the $f_{7/2}$ shell are
shown for \up48Cr and \up44Ti. } \label{fig-Cr48SPConf}
\end{figure}

Equilibrium deformations calculated in the CNS model at different
spin values are shown in Fig.\ \ref{fig-EE-EqConf}. The quadrupole
deformation of the ground state, $\varepsilon \approx 0.23$,
implies a fairly large mixing of the $f_{7/2}$ and $p_{3/2}$
shells, see the Nilsson diagram in Fig.\ \ref{fig-Cr48SPConf}.
Since all four positive $m$-states of the $f_{7/2}$ subshell are
occupied for protons as well as for neutrons in the
band-terminating state 16$^+$ (cf.\ Fig.\ \ref{fig-Cr48Occ}), the
nucleus obtains a spherical shape. A gradual change in
deformation, as the spin increases from 0\parp\ in the ground
state to the 16$^+$ state (Fig.\ \ref{fig-EE-EqConf}), explains
the main changes in the occupation numbers of the $f_{7/2}$ and
$p_{3/2}$ shells, seen in Fig.\ \ref{fig-Cr48Occ}. When the
deformation decreases, the $j$-shells get less mixed.

%----------
%----------

%----------
%----------
%----------
%----------

%----------

\subsubsection{Negative-parity band in $^{48}$Cr}
\label{sssect-Cr48NegPar}
A negative-parity band in the $pf$ shell
nuclei can be obtained by exciting one particle (a proton or a
neutron) from a $d_{3/2}$ orbit to an unoccupied $f_{7/2}$ orbit,
forming a configuration $d_{3/2}^{-1}f_{7/2}^9$. Denoting the
signature quantum number by $\alpha,$ we can write the even-spin
configurations (with $\alpha_{\rm tot}=0$) as
$[\alpha(d_{3/2}^{-1}), \alpha(f_{7/2}^{\rm odd})]=[-1/2,1/2]$ and
$[1/2,-1/2]$, while the odd-spin configurations (with $\alpha_{\rm
tot}=1$) are $[-1/2,-1/2]$ and $[1/2,1/2]$. In the CNS
calculations, the odd-spin band with $K^\pi = 4^-$ (excitation
$d_{3/2,3/2}^{-1} f_{5/2,5/2}^{}$) is clearly energetically favored.
Other configurations have higher energies. The two bands with
$\alpha_{\rm tot} = 0$ have similar energy values but their
equilibrium deformations are somewhat different.

\begin{figure}[tbp]
\includegraphics*[width=7.5cm, angle=0]{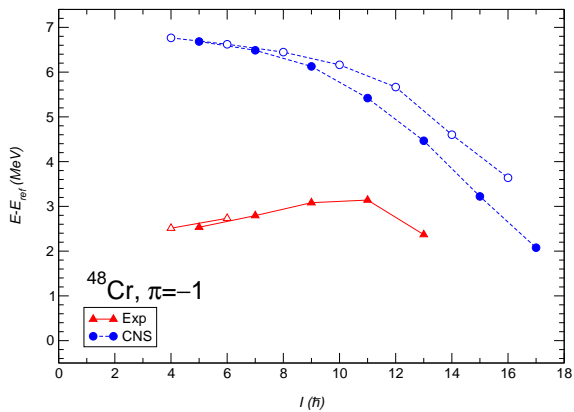}
\caption{ (Color online) Energies of the negative-parity band in
\up48Cr: experimental data \cite{Cr48Exp} are shown using
triangles, and the CNS results are shown using circles. Filled and
empty symbols are used to differentiate between the $\alpha=1$ and
$\alpha=0$ bands. The CNS energies are normalized in the same way
as the ground-state band in Fig.\ \ref{fig-Cr48ERef}. }
\label{fig-Cr48NP}
\end{figure}

Measured and calculated energies of the negative-parity band in
\up48Cr are shown in Fig.\ \ref{fig-Cr48NP}. The measured
negative-parity band starts with a 4\parm\ state at $E_x=3.53$ MeV
\cite{Lenzi02a,Ur04}. Unpaired CNS calculation gives the
excitation energy of this state $E_x(4^-)=3.65$ MeV, which is
unexpectedly close to the experimental value. To calculate the
unnatural parity bands within the shell model, the model space
must be expanded to include at least the $d_{3/2}$ shell. This
larger model space calculation is outside of the scope of our
paper. We refer to Ref.\ \cite{Cr48Exp} where such a calculation
was reported. Assuming the excitation of a $d_{3/2}$ nucleon to
the $pf$ shell, the lower part of the spectrum could be well
described, while higher spins were described poorly. These
energies are not included in Fig.\ \ref{fig-Cr48NP}.

Two CNS bands are shown in Fig.\ \ref{fig-Cr48NP}: the favored
configuration with $\alpha_{\rm tot}=1$ and one of the
$\alpha_{\rm tot}=0$ configurations. In both bands the signature
of the $d_{3/2}$ hole is $\alpha=-1/2$. The CNS calculation
predicts a somewhat bigger moment of inertia than observed
experimentally. This can be related to the absence of pairing
correlations in the model. If we assume that the spin dependence
of the pairing energy is similar to that in \up49Cr (see Fig.\
\ref{fig-Cr49Epair}), an approximate prediction of the unobserved
energies in the negative-parity band may be obtained (not shown in
the figure). In particular, the contribution from the pairing
correlations is expected to approximately double the signature
splitting predicted by the CNS calculations shown in
Fig.\ \ref{fig-Cr48NP}.

%-------------
%-------------

\begin{figure}[tbp]
\includegraphics*[width=7.5cm, angle=0]{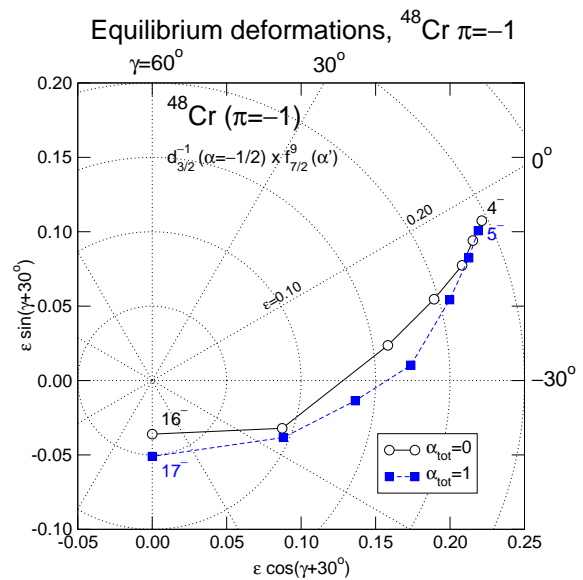}
\caption{ (Color online) CNS calculated equilibrium deformations
in the \up48Cr negative-parity band. The odd particle in $f_{7/2}$
has $\alpha'=+1/2$ for the even-spin states ($\alpha_{\rm tot}=0$)
and $\alpha'=-1/2$ for the odd spins ($\alpha_{\rm tot}=1$). }
\label{fig-Cr48EqDefNegPar}
\end{figure}

The calculated equilibrium deformations for the negative-parity
bands are shown in Fig.\ \ref{fig-Cr48EqDefNegPar}. The 4\parm\
band-head has a similar deformation as the 0\parp\ ground state,
in agreement with the measured $B(E2)$ values
\cite{A48NDS,Cr48Exp}. However, already this state has a sizeable
value of the triaxiality parameter, $\gamma$. The non-axiality
gets stronger as spin increases, and the band terminates in a
non-collective prolate shape, having $\gamma=-120^\circ$. In
fact, this negative-parity band in \up48Cr exhibits the largest
calculated negative-$\gamma$ deformation in the region of $pf$
shell nuclei. This can be understood from Fig.\ \ref{fig-SPOrbits}
below, where we present the relevant single-particle routhians. A
hole in the $N=2$, $\alpha = +1/2$ orbital, shown in the figure by
a solid line, will have a strong polarization effect towards
negative-$\gamma$ values. The excited particle will occupy the
25th or 26th orbital, which are both essentially
$\gamma$-independent. Thus, the net effect is a large
negative-$\gamma$ deformation of the bands. Since at high spins
the spectroscopic quadrupole moment is proportional to
$\sin(\gamma+30^\circ)$ \cite{Ham83}, we expect it to be small in
this band. The stretched $B(E2)$ values are not much affected by
this kind of triaxiality (i.e., by negative $\gamma$). Thus a
decrease in collectivity towards the band-terminating state is
expected.

%----------
%----------

% -------------------------------------
\subsubsection{The nuclei \up44Ti and \up46Ti}
\label{sssTi}

\begin{figure}[tbp]
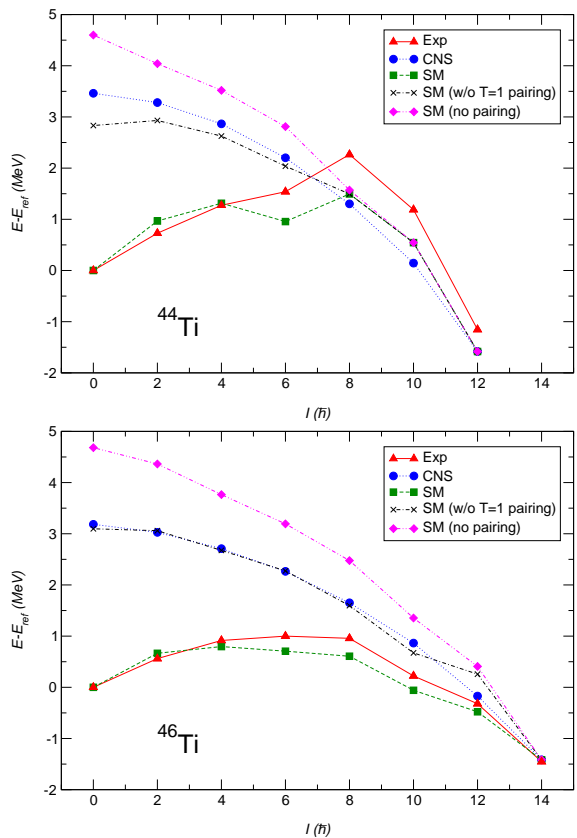

\includegraphics*[width=7.5cm, angle=0]{fig-Ti44Eref.eps}
\includegraphics*[width=7.5cm, angle=0]{fig-Ti46Eref.eps}
\caption{ (Color online)
Excitation energies of the yrast states in \up44Ti and \up46Ti plotted
relative to a rotational reference. The experimental energies in
\up44Ti and \up46Ti are taken from Refs.\ \cite{Ti44Exp} and
\cite{Bucurescu03}, respectively.
} \label{fig-TiERef}
\end{figure}

Calculated and experimental energies of the yrast states in
\up44Ti and \up46Ti are compared in Fig.\ \ref{fig-TiERef}. These
two nuclei have several measured bands, see Refs.\ \cite{Ti44Exp} and
\cite{Bucurescu03} respectively, but we restrict ourselves to the
yrast bands only. As expected from the neglect of pairing, the CNS
energies deviate from experimental data at low spins. It might
appear surprising that the agreement between the shell model
results and experiment is not outstanding for \up44Ti. However, it
is not unexpected, since the $pf$-shell model space is rather
restricted for this small number of particles, and the excitations
out of the $sd$ shell play an important role in this nucleus
\cite{Ti44Exp}.

%----------
%----------

As we already discussed in connection with \up48Cr, it is more
relevant to compare the CNS energies with the shell model results
obtained when the pairing force is removed from the interaction.
Therefore Fig.\ \ref{fig-TiERef} shows two additional lines: the
calculated energies when the $T=1$ pairing is subtracted, and when
both the $T=1$ and $T=0$ pairings are subtracted. Pairing does
not contribute much to the energy for the states above $I=6$. This
is particularly true for \up44Ti, where both the $T=0$ and $T=1$
pairing energy contributions are approximately zero at $I>6$.
The CNS energies compare well with those unpaired
calculations, especially with the one where only the $T=1$ pairing
is removed. Furthermore, the $T=1$ pairing is the main cause for
backbending at $I \approx 10$ in both nuclei, since the $T=0$
pairing has a smooth dependence, decreasing with spin. All these
features are similar to those previously discussed for \up48Cr,
see Fig.\ \ref{fig-Cr48ERef}.

%----------
\begin{figure}[tbp]
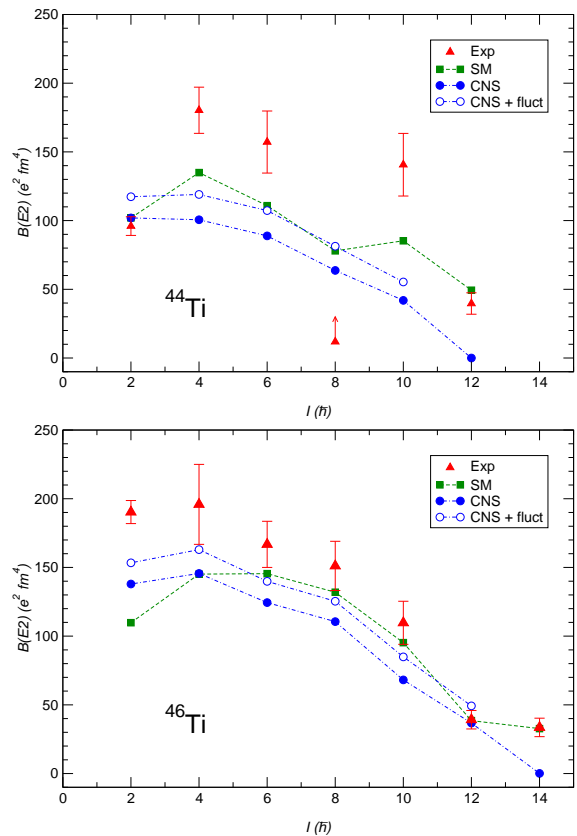

\includegraphics*[width=7.5cm, angle=0]{fig-Ti44BE2.eps}
\includegraphics*[width=7.5cm, angle=0]{fig-Ti46BE2.eps}
\caption{ (Color online)
The strength of the stretched $E2$ transitions along the yrast bands
of \up44Ti and \up46Ti. Measured values are taken from Refs.\
\cite{Ti44BE2,Schielke03} and \cite{Brandolini04}, respectively.
}
\label{fig-TiBE2}
\end{figure}
%----------

Having half-filled $f_{7/2}$ shells for both protons and neutrons,
the nucleus \up48Cr is the most collective nucleus in the region.
By removing one pair of protons from \up48Cr we arrive at the
nucleus \up46Ti. If also a pair of neutrons is removed, we get
\up44Ti. The quadrupole deformations $\varepsilon$=0.15, 0.19 and
0.23 are predicted for the ground states of the nuclei \up44Ti,
\up46Ti, and \up48Cr, respectively, as illustrated in Fig.\
\ref{fig-EE-EqConf}. The origin of this deformation increase with
the number of valence particles can be understood by analyzing the
Nilsson diagram, see Fig.\ \ref{fig-Cr48SPConf}. The lower
$f_{7/2}$ orbitals are deformation-driving, thus a gradual
increase in the collectivity is expected with the increasing
number of valence particles. This was indeed observed in the
decays of the 2\parp\ states to the ground states via $E2$
transitions. The $B(E2)$ values in these three nuclei are: 96
\cite{Schielke03}, 190 \cite{Brandolini04}, and 311 $e^2\,{\rm
fm}^4$ \cite{A48NDS}. With 4, 6 and 8 valence particles in the
$f_{7/2}$ shell, the rotational bands in \up44Ti, \up46Ti and
\up48Cr terminate at $I$= 12$^+$, 14$^+$ and 16$^+$, respectively.

Available experimental information and the calculated $B(E2)$ values
for the two titanium isotopes are shown in Fig.\ \ref{fig-TiBE2}. The
measured values in \up44Ti are taken from Ref.\ \cite{Ti44BE2}. The
lifetimes of two levels, 2\parp\ and 4\parp, were remeasured recently
by Schielke {\em et al\/} \cite{Schielke03}.  Using the formula
$\tau_{1/2}=\log_e(2)/(1.22\times 10^9 E_\gamma^5 B(E2))$, from the
values of 3.97(28)ps and 0.65(6)ps we deduce the $B(E2)$ strengths
$96.0\pm6.8$ $e^2\,{\rm fm}^4$ and $180.3\pm16.8$ $e^2\,{\rm fm}^4$,
respectively.  Experimental information on \up46Ti is taken from Ref.\
\cite{Brandolini04}.

Both the shell model and the CNS model suggest a decrease in
collectivity as the angular momentum increases in \up46Ti and
\up44Ti, in a similar way as was found in \up48Cr
\cite{Cr48triax}. It is astonishing that the transition strengths
are so similar in the two models, especially if the CNS results
are corrected for quantum fluctuations around the equilibrium, as
was described in Ref.\ \cite{Cr48triax}. The yrast bands
terminate at $I=12^+$ and 14$^+$ with some remaining collectivity:
According to experiment, the $E2$ transitions from these states
have $B(E2)$ values of the order of 4 W.u.

% -------------------------------------
\subsection{Odd-even nuclei}
\label{ssect-OENuclei} In this subsection we discuss odd-even mass
nuclei with $N=Z\pm1$. In particular, we shall focus on the
signature splitting of the yrast band. In subsection
\ref{sssect-A4547} we present results for the nuclei with the mass
$A=45$ and 47, while results for \up49Cr are presented in
subsection \ref{sssect-Cr49}. The role of pairing for the
backbending as well as signature splitting is discussed in
subsection \ref{sssect-Pairing}.

% ---------------------
\subsubsection{$A$=45 and $A$=47 nuclei}
\label{sssect-A4547}

The mirror nuclei,
$_{22}^{45}$Ti$_{23}^{}$-$_{23}^{45}$V$_{22}^{}$ and
$_{24}^{47}$Cr$_{23}^{}$-$_{23}^{47}$V$_{24}^{}$, have identical
spectra in the shell model description if the Coulomb and other
isospin-nonconserving interactions are neglected in favor of a
good isospin. In CNS, where protons and neutrons have different
single-particle spectra due to the Coulomb interaction, the
predicted properties of mirror nuclei are still very similar,
since the effect of the Coulomb force (and other isospin-violating
interactions) is relatively small in this mass region, i.e. the
experimental mirror energy difference is below 100 keV
\cite{Garrett01,Zuker02,Tonev02}. Because of this similarity in
spectra, we present the calculated results only for \up45Ti and
\up47V. Similarly, like in the case of even-even nuclei, we
restrict our discussion to the yrast bands.

\begin{figure}[tbp]
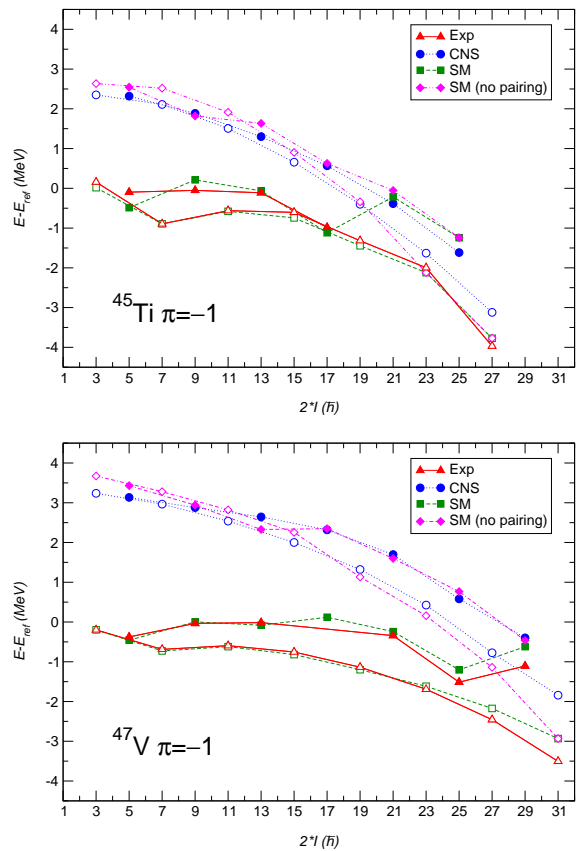

\includegraphics*[width=7.5cm, angle=0]{fig-Ti45Eref.eps}\\[3mm]
\includegraphics*[width=7.5cm, angle=0]{fig-V47Eref.eps}
\caption{ (Color online)
Energies of yrast states in \up45Ti and
\up47V. Experimental data is taken from Refs.\ 
\cite{Bednarczyk98} and \cite{Brandolini01v47}, respectively.
The levels $21/2^-$ and $25/2^-$ in \up45Ti, and
the 17/2\parm\ state in \up47V are not known.
Open and empty symbols are used to distinguish the signature
$\alpha=+1/2$ and $\alpha=-1/2$ bands.
} \label{fig-45V47CrERef}
\end{figure}

Experimental and calculated energies of the yrast states in \up45Ti
and \up47V are compared in Fig.\ \ref{fig-45V47CrERef}. In general,
the shell model calculations describe the measured energies
well. Experimental values are taken from Refs.\ \cite{Bednarczyk98}
and \cite{Brandolini01v47}, respectively. The $\alpha =+1/2$ levels in
\up45Ti are only known up to $I^\pi = 17/2^-$, and the energy of the
17/2\parm\ level in \up47V is unknown. In the mirror \up47Cr nucleus,
the 17/2\parm\ level was measured at 3.77 MeV \cite{Cameron94}. These
missing levels inhibit a discussion of the pair-alignment process in
the backbending region, since the mirror energy difference cannot be
extracted.

The ground states of \up45Ti and \up47V are described in the CNS with
an odd neutron and proton, occupying the $\Omega=3/2$ state of the
$f_{7/2}$ shell, see Fig.\ \ref{fig-Cr48SPConf}. This gives rise to a
fairly large rotation-induced calculated signature splitting which
varies smoothly with spin. The different character of the
signature partners, observed experimentally in both nuclei, is thus
expected to originate from the pairing force. This conclusion is
supported by a remarkably good agreement between the unpaired shell
model and the CNS energies, see Fig.\ \ref{fig-45V47CrERef}. Not only
the moment of inertia comes out correct but also the signature
splitting is fairly well described.  Since a similar behavior is also
seen in \up49Cr (as we will present in the next subsection), we
discuss the role of pairing for the signature splitting in subsection
\ref{sssect-Pairing}.

%----------
%----------

%----------
\begin{figure}[tbp]
\includegraphics*[width=7.5cm, angle=0]{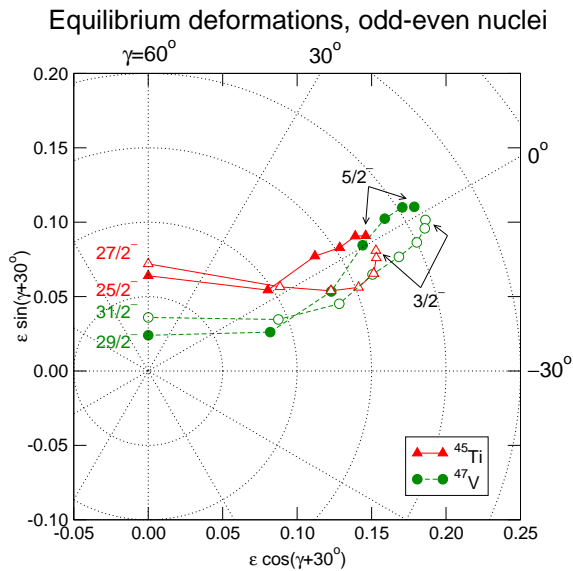}
\caption{(Color online) Equilibrium deformations along the yrast
band in odd-even nuclei \up45Ti (\up45V) and \up47V (\up47Cr)
calculated in the CNS. } \label{fig-OE-EqConf}
\end{figure}
%----------

%--------
\begin{figure}[tbp]
\begin{center}
\includegraphics*[width=8cm, angle=0]{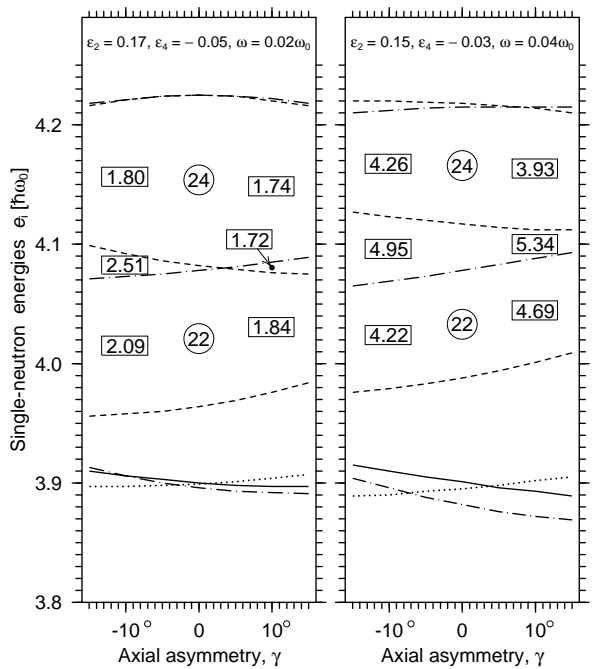}
\end{center}
\caption{ Dependence of the single-particle routhians on the
non-axiality parameter $\gamma$ at deformations and rotational
frequencies relevant for $I \approx 4$ (left panel) and $I \approx
8$ (right panel) states in $^{45}$Ti. The encircled numbers indicate the
number of orbitals below that point. The boxed numbers show the
spin contributions from all 22, 23, and 24 neutrons at $\gamma\pm
10^{\circ}$. Positive-parity orbitals from the $sd$ ($N=2$) shell
are drawn by full ($\alpha=+1/2$) and dotted ($\alpha=-1/2$)
lines. Other lines are negative-parity $pf$-shell orbitals:
$\alpha=+1/2$ orbitals are drawn with dashed lines, while
$\alpha=-1/2$ orbitals are drawn with dash-dotted lines. The
diagram is drawn for neutrons but the proton orbitals have very
similar properties. } \label{fig-SPOrbits}
\end{figure}
%----------

Calculated equilibrium deformations for the two signature-partner
bands in \up45Ti and \up47V are shown in Fig.\ \ref{fig-OE-EqConf}. An
interesting staggering of the shape is seen. The $\alpha=+1/2$ band
has positive $\gamma$ values, while its signature-partner,
$\alpha=-1/2$ band has negative $\gamma$ values. The different
$\gamma$-deformations of the two signature bands should mainly affect
the quadrupole properties, while energies are less sensitive to the
small changes of the triaxiality parameter. The equilibrium
deformations change smoothly from $\varepsilon=0.17$ (0.21) in the
ground state of \up45Ti (\up47V) to $\varepsilon\approx0.07$ (0.03) in
the band-terminating state, having non-collective oblate shape
($\gamma=60^\circ$). The values of $\gamma$ parameter do not exceed
$10^\circ$ for most of the states prior to termination.

This different $\gamma$-preference may be understood from the
single-particle routhians plotted at fixed rotational frequencies
in Fig.\ \ref{fig-SPOrbits}. The encircled numbers show the
number of particles below that point. One can clearly see that
the 23rd single-particle orbital prefers either a positive or
negative $\gamma$ value, depending on whether the $\alpha=+1/2$ or
$-1/2$ branch is occupied. This single-particle orbital is crucial
for both $_{22}^{45}$Ti$_{23}^{}$ and $_{24}^{47}$Cr$_{23}^{}$,
because neither the 22 protons in $^{45}$Ti nor the 24 neutrons in
$^{47}$V have any strong preference in $\gamma$. The weak
$\gamma$-dependence for particle number 24 is also seen from the
fact that the ground band of $^{48}$Cr is calculated to be nearly
axially deformed at low spin values \cite{Cr48triax}. Since the
single-neutron and single-proton level schemes are very similar,
the same arguments apply for the nuclei $_{23}^{45}$V$_{22}^{}$
and $_{23}^{47}$V$_{24}^{}$.

The routhians of Fig.\ \ref{fig-SPOrbits} can also be used to
illustrate the triaxial properties of nuclei with $N$ or $Z$ equal
to either 22 or 24. The summed effect of the two lowest $f_{7/2}$
orbitals (dashed and dash-dotted lines below particle number 22 in
Fig.\ \ref{fig-SPOrbits}) gives no strong preference in the
$\gamma$-direction. The 23rd and 24th orbitals have forces driving
into different $\gamma$-directions, leading to opposite
$\gamma$-deformations. Their added-up effect cancels the
$\gamma$-sign preference for 24 particles. There is also a pretty
strong driving force of the highest $N=2$, $\alpha = +1/2$
orbital, which explains large negative-$\gamma$ deformations in
the 1p-1h bands, as we discussed in the case of \up48Cr,
see~\ref{sssect-Cr48NegPar}.

% -------------------------------------
\subsubsection{The nucleus \up49Cr}
\label{sssect-Cr49}

%----------
\begin{figure}[tbp]
\includegraphics*[width=7.5cm, angle=0]{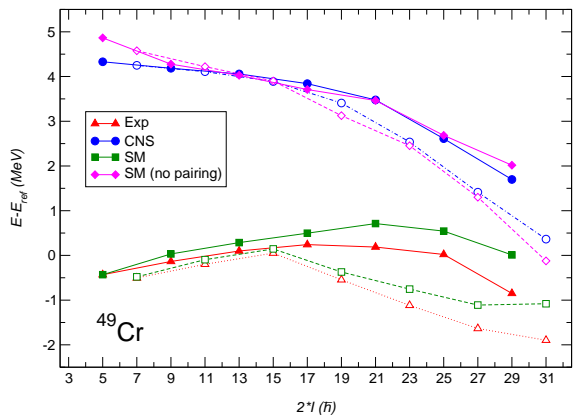}
\caption{ (Color online) Experimental \cite{Brandolini01v47} and
calculated energies of yrast states in \up49Cr. In addition to the
CNS and the shell model calculations (circles and squares,
respectively), shell model energies without pairing are shown
(diamonds). } \label{fig-Cr49ERef}
\end{figure}
%----------

%----------
\begin{figure}[tbp]
\includegraphics*[width=7.5cm, angle=0]{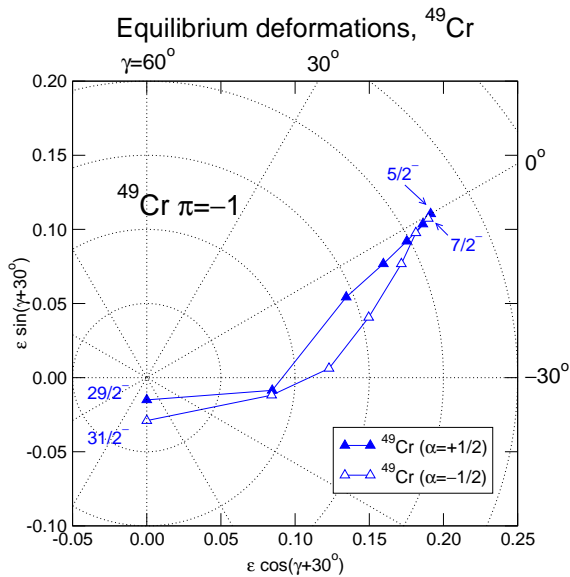}
\caption{ (Color online)
The CNS predicted equilibrium deformations
along the yrast band in \up49Cr.
} \label{fig-Cr49EqDef}
\end{figure}
%----------

%----------
\begin{figure}[tbp]
\includegraphics*[width=7.5cm, angle=0]{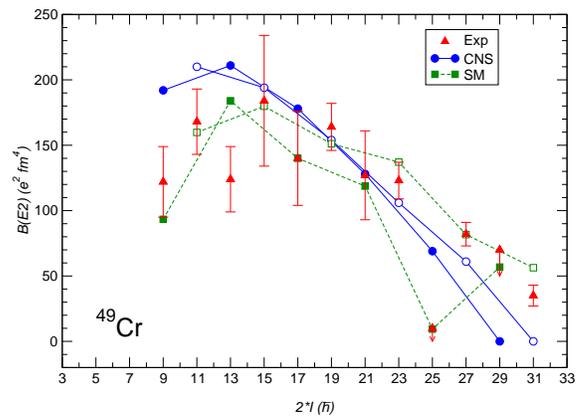}\\[2mm]
\includegraphics*[width=7.5cm, angle=0]{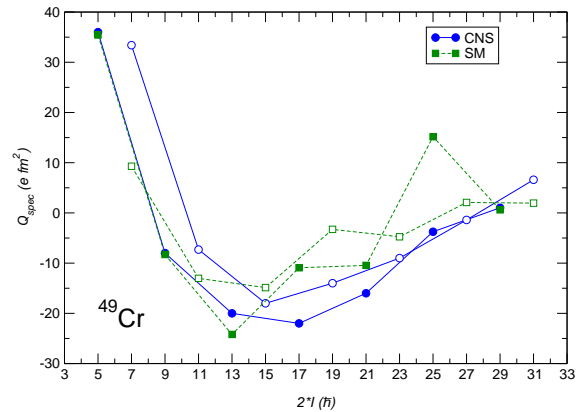}
\caption{ (Color online)
Quadrupole properties of yrast states in \up49Cr. The experimental
$B(E2)$ values are taken from Ref.\ \cite{Brandolini01v47}.
}
\label{fig-Cr49BE2}
\end{figure}
%----------

Calculated and experimental yrast energies for \up49Cr are shown
in Fig.\ \ref{fig-Cr49ERef}. The observed splitting of the
signature partners is small at low spins. However, at the spin
value $I=15/2$ the $\alpha=-1/2$ band suddenly changes its smooth
behavior in a backbending. The other signature band depends on
angular momentum in a smoother way up to the band termination at
31/2\parm. This behavior of the two signature-partner bands is
well reproduced in the shell model calculations \cite{A47A49SM},
although there is a systematic deviation at higher spins. The CNS
calculations show large deviations, particularly for low spins, as
is expected due to the lack of pairing correlations in this model.
From the Nilsson diagram (Fig.\ \ref{fig-Cr48SPConf}) one can see
that in the ground-state configuration the odd neutron occupies
the $\Omega$=5/2 orbital of the $f_{7/2}$ shell. This large
$\Omega$ value causes a rather small signature splitting in the
CNS calculation. Similarly as in the case of \up45Ti and \up47V,
the shell model signature splitting reduces if the pairing
interaction is removed. These values agree very well with the CNS
calculation. We explore this observation in a greater detail in the
next subsection.

Calculated equilibrium deformations in the lowest $\alpha=+1/2$ and
$-1/2$ bands in \up49Cr are shown in Fig.\ \ref{fig-Cr49EqDef}. The
spin dependence of the quadrupole deformation parameter $\varepsilon$
was briefly discussed in Ref.\ \cite{Bentley98}. Here we present a
more detailed study.  Both signature bands prefer negative $\gamma$
values already at low spins. They terminate in non-collective states
with $\gamma=-120^{\circ}$ at spins $I^\pi$=29/2$^-$ and 31/2$^-$,
respectively. This value of the asymmetry parameter corresponds to a
prolate nucleus with the angular momentum aligned along the symmetry
axis.

Based on the equilibrium deformations, $B(E2)$ values and
spectroscopic moments $Q_{\rm spec}$ were calculated. They are
shown in Fig.\ \ref{fig-Cr49BE2} together with experimental data
on $B(E2)$ \cite{Brandolini01v47} and the shell model results. (No
experimental information on spectroscopic quadrupole moments is
available.) The agreement between the two model predictions is
remarkable. As seen in Fig.\ \ref{fig-Cr49EqDef}, the nuclear
shape gradually changes along the band from axially-symmetric
prolate to triaxial shapes with negative $\gamma$ values.
Simultaneously, the quadrupole parameter $\varepsilon$ gets
smaller, which explains the gradual decrease of the $B(E2)$ values
for both signatures. It is interesting to note that the quadrupole
properties along the signature bands are the same, although the
energies in one of them shows a backbending behavior.

\begin{figure}[tbp]
\includegraphics*[width=7.5cm, angle=0]{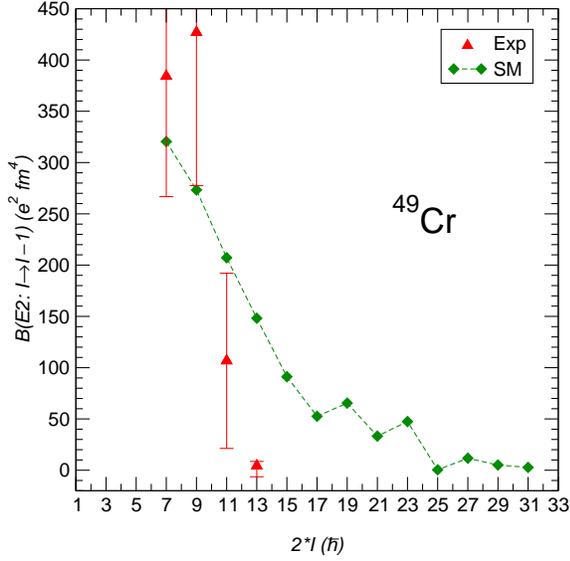}
\caption{ (Color online)
Unstretched $B(E2)$ values in \up49Cr: experimental data
\cite{A49NDS} and the shell model predictions \cite{Brandolini01v47}.
The staggering occurring at higher spins indicates non-axiality
of the nuclear shape \cite{Ham83}.
}
\label{fig-DeltaI1}
\end{figure}

As discussed above, the shapes of both signatures in \up49Cr
change with increasing spin and become clearly triaxial after spin
$I>17/2$, where the rotation takes place around the intermediate
axis (negative value of $\gamma$ in the Lund convention). A
signature of nuclear triaxiality in an odd-mass nucleus is the
staggering of $B(E2,\Delta I = 1)$ values, as discussed by
Hamamoto and Mottelson \cite{Ham83}. Thus, we plotted them
\cite{A49NDS,Brandolini01v47} in Fig.\ \ref{fig-DeltaI1}. It is
clearly seen that the staggering in the shell model values of
$B(E2; \Delta I=1)$ indeed appears above the angular momentum
$I=17/2$, when the shapes calculated in CNS become triaxial (Fig.\
\ref{fig-Cr49EqDef}).

%----------
%----------

% -----------------------------------------
\subsubsection{Pairing, Backbending and Signature Splitting}
\label{sssect-Pairing}

% ----------------------------------------
\begin{figure}[tbp]
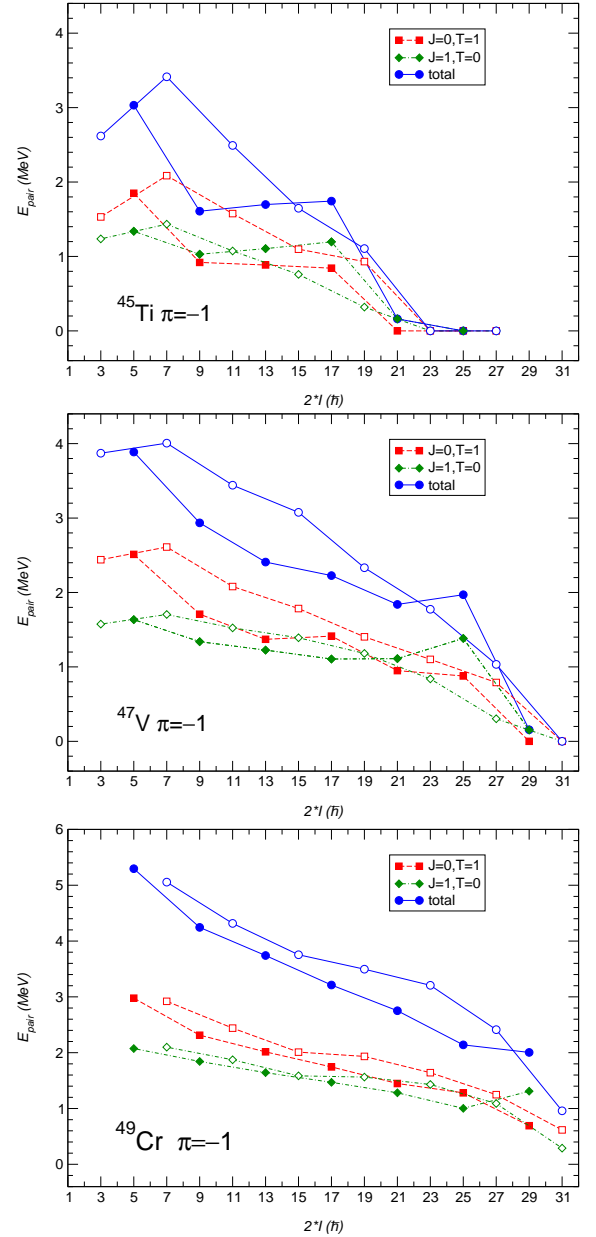

\includegraphics*[width=7.5cm, angle=0]{fig-Ti45Epair.eps}\\
\includegraphics*[width=7.5cm, angle=0]{fig-V47Epair.eps}\\
\includegraphics*[width=7.5cm, angle=0]{fig-Cr49Epair.eps}
\caption{ (Color online)
Pairing energy in nuclei \up45Ti, \up47V, and \up49Cr.
Filled and open symbols are used to distinguish different
signature bands: $\alpha=+1/2$ and $-1/2$, respectively.
}
\label{fig-Ti45V47Epair}
\label{fig-Cr49Epair}
\end{figure}
% ----------------------------------------

%----------
\begin{figure}[tbp]
\includegraphics*[width=7.5cm, angle=0]{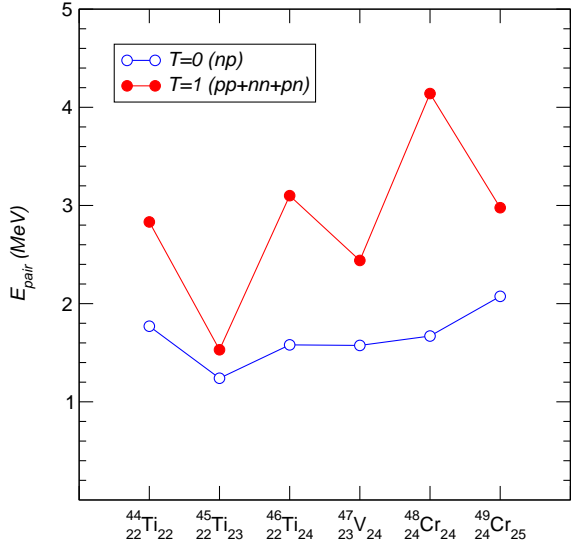}
\caption{ (Color online) The shell model calculated ground-state
pairing energy. The plotted $T=0$ pairing energy in even-even nuclei 
is calculated as a difference between the total pairing energy and
the $T=1$ pairing energy. } \label{fig-GS-Pairing}
\end{figure}
%----------

%----------
\begin{figure}[tbp]
\includegraphics*[width=7.5cm, angle=0]{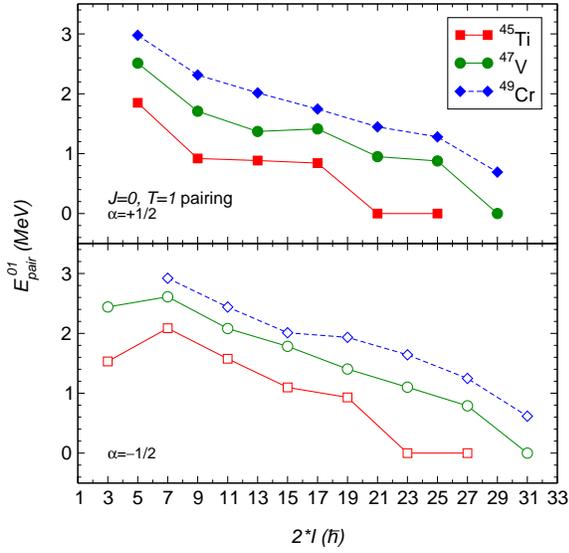}
\caption{ (Color online) The $J=0,\,T=1$ pairing energy calculated
in the shell model for the odd-$A$ nuclei
 \up45Ti, \up47V and \up49Cr. The upper (lower) panel shows contributions to
 the $\alpha=+1/2$ ($-1/2$) signature bands.
} \label{fig-T1Pairing-Actual}
\end{figure}
%----------

The pairing energy calculated in the shell model has been
extensively used above. Subtracting this energy from the shell
model value, an unpaired shell model energy could be obtained.
Remarkably, this agreed very well with the energies calculated
within the unpaired CNS model. This leads us to a more detailed
study of the shell model pairing energy, its role in causing
backbending of the ground-state band, and its contribution to the
signature splitting.

In Fig.\ \ref{fig-Cr49Epair} we show separately the $T=0$, the $T=1$
and the sum of both pairing energies as a function of spin for the
three odd-even nuclei \up45Ti, \up47V, and \up49Cr. The $T=0$ pairing
energy behaves smoothly in a similar way as it does in the even-even
nucleus \up48Cr (see Fig.\ 4 in \cite{Cr48pair}). Its contribution
decreases from 1-2 MeV in the ground state to a small or zero
contribution in the spin-aligned state (the highest angular momentum
state shown in the figure). The $T=1$ pairing shows a pronounced
odd-even effect, see Fig.\ \ref{fig-GS-Pairing}. The odd nucleon in
the three studied odd-even nuclei implies a blocking effect that
weakens the isovector pairing energy by about 1~MeV in the ground
state, as compared to the neighboring even-even nuclei. The isoscalar
pairing energy has a much smoother mass dependence. Note that the
isoscalar pairing contributions seem to follow the same trend along
the yrast band as the isovector pairing (Fig.\ \ref{fig-Cr49Epair}). 
The latter is easy to understand in terms of the seniority. To
highlight the change of the $T=1$ pairing energies, we plot them
separately in Fig.\ \ref{fig-T1Pairing-Actual}. One can notice that
the increasing number of particles increases the collectivity
(configuration mixing), and diminishes the irregularities in the spin
dependence of the pairing energy.

% ------- backbending

%-------------
\begin{figure}[tbp]
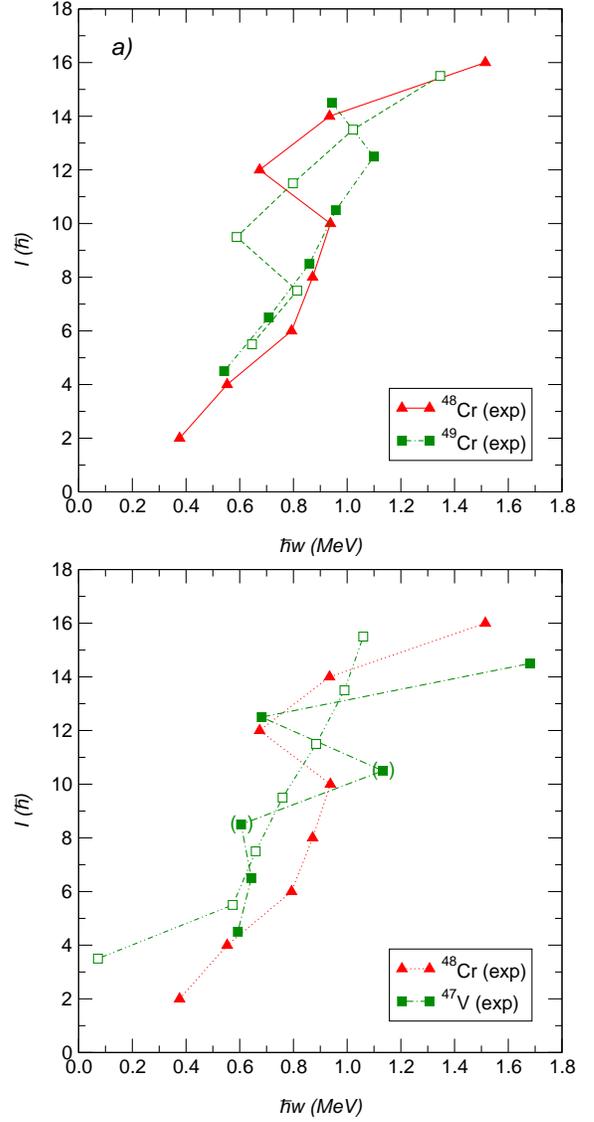

\includegraphics*[width=7.5cm, angle=0]{fig-CrIsoBackbend_a.eps}\\
\includegraphics*[width=7.5cm, angle=0]{fig-CrIsoBackbend_b.eps}
\caption{ (Color online) Spin versus rotational frequency in
$_{}^{48,}{}_{24}^{49}$Cr (top, $a$) and \updown4723V, \updown4824Cr
(bottom, $b$). Experimental data are used for Cr isotopes, while for
the unknown 17/2\parm\ level in \up47V, the energy of this level in
the mirror nucleus \up47Cr is assumed, and the affected transitions
are marked by parentheses. \label{fig-Cr49Bkbend} }
\end{figure}
%-------------

As we shall see, the changes of the pairing energy with increasing
spin causes backbending. Backbending of the yrast band in the nucleus
\up48Cr has received a great deal of attention, and the usual
explanation in terms of a band-crossing is only partly supported by
the experimental data. Another important suggestion to explain the
observed behavior \cite{Brandolini04} is the dominance of the
seniority $v=4$ configuration at spin 12\parp\ \cite{Cr48triax}. 
Studies of the spectra of the neighboring nuclei with one proton or
neutron either added or removed show energy irregularities along the
yrast band. However, it seems to be overseen that this irregularity
occurs only in one signature-partner, and not in the other. This is
seen from Fig.\ \ref{fig-Cr49Bkbend}, where we show the total angular
momentum as a function of the rotational frequency
$\hbar\omega=[E_\gamma(I)-E_\gamma(I-2)]/2$. Cameron et al
\cite{Cameron90} showed the two signatures in \up49Cr, but incorrect
spin assignments of high-spin states led to a wrong similarity between
the bands. Mart{\'\i}nez et al \cite{A47A49SM} stated that the
backbending behavior would be seen, if the energies were plotted as
spin versus the rotational frequency.  However, they did not discuss
that in a greater detail nor mention the different behavior of the
signature bands.

In a backbending plot, Fig.\ \ref{fig-Cr49Bkbend}$a$, we compare
the behavior of the two signature bands in \up49Cr with the yrast
band in \up48Cr. The lowest signature band in \up49Cr with
$\alpha=+1/2$ has a very similar behavior as the yrast band in
\up48Cr with the difference that the backbend occurs at a lower
spin value, $I=19/2$. The other signature band has a different
spin dependence on frequency and exhibits no backbending for spin
values $I \approx 10$. The change in the last transition is
rather an effect of a band termination \cite{Cr48BT} than a
backbend. A similar difference in the signature partners is also
observed in \up47V, see Fig.\ \ref{fig-Cr49Bkbend}$b$. A
difference from \up49Cr is that the backbending occurs in the
other signature band, $\alpha=+1/2$. Due to the unmeasured
17/2\parm\ state in \up47V, only the sum of the two transitions
$21/2^-\rightarrow17/2^-$ and $17/2^-\rightarrow13/2^-$ is known.
If, however, we assume that the excitation energy of this state is
close to the value measured in the mirror nucleus \up47Cr (as we
did in the figure), the backbend is clearly seen. Note that
independently of the exact value of the $17/2^-$ energy this band
will show a backbending.

% ----------- signature, backbend, pairing

As we saw in previous subsections, the sum of the $T=0$ and $T=1$
pairing energies calculated in the shell model and shown in Fig.\
\ref{fig-Cr49Epair} seems to describe the missing pairing energy in
the CNS calculation for odd-$A$ nuclei quite well, see Figs.\
\ref{fig-45V47CrERef} and \ref{fig-Cr49ERef}. When the pairing energy
is subtracted from the shell model energies, no backbending is seen,
and the two signature bands behave in a similar way. This implies that
pairing causes backbending. Furthermore, pairing increases the
signature splitting which is already present because of a rotational
coupling. The latter is well described in the CNS model. Further we
shall discuss the pairing contribution to the signature splitting that
also reveals the different backbending behavior in the two signature
bands.

In a simple $f_{7/2}$ shell model for the odd-even nuclei \up45Ti,
\up47V, and \up49Cr with only $T=1$ pairing force considered, the
pairing energy in both signature bands is the same. The pairing
energy contribution is, however, shifted by 1 unit of angular
momentum. For example, the amount of pairing in 5/2\parm\ and
7/2\parm\ states in \up49Cr is the same, because there is no
difference in the seniorities of protons and neutrons in these two
states: $v_p=0$, $v_n=1$. To gain angular momentum, one needs to
break pairs, and it is energetically favored to do it in a similar
way in both signature partners. Since the $f_{7/2}$ shell is
dominant, this similarity is seen in the pairing energy of states
with the spins $I$ in the signature $\alpha=+1/2$ band and $(I+1)$
in the signature partner, see Fig.\ \ref{fig-T1Pairing-Actual}.
The curve that describes the pairing energy contribution to the
$\alpha=-1/2$ band is approximately the same as for the
$\alpha=+1/2$ band, but is moved to the right by one unit of
angular momentum. This gives a contribution to the signature
splitting, since it is defined as the energy difference between
the two signature partner bands calculated at a fixed value of
angular momentum.

%----------
\begin{figure}[tbp]
\includegraphics*[width=7.5cm, angle=0]{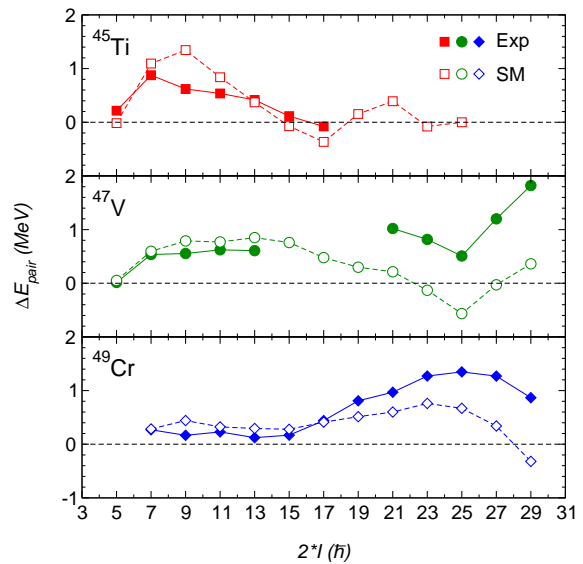}
\caption{ (Color online) The signature splitting from the total
pairing contributions (open symbols) and the experimental values
(filled symbols) in \up45Ti, \up47V and \up49Cr. }
\label{fig-PairingSignatureSplit}
\end{figure}
%----------

The contribution from $T$=0 and $T$=1 pairing to the signature
splitting, $\Delta E_{pair}$, is studied in Fig.\
\ref{fig-PairingSignatureSplit} for the considered odd-even
nuclei. We also show the experimental signature splitting that is
seen to follow the trends of the shell model calculation. The
difference between the experimental signature splitting and the
pairing signature splitting comes from rotational coupling, which
is well described in unpaired CNS calculations. The signature
splitting from the $T=0$ pairing behaves smoothly in a similar way
as in the even-even nucleus $^{48}$Cr (see Fig.\ 4 in
\cite{Cr48pair}), while the irregular behavior comes from the
$T=1$ part. This can be explained to some extent by a trivial
shift by one unit of angular momentum between the two signature
partners, as was discussed above. The change of slope of $\Delta
E_{pair}$ around $I=25/2$ seen in $^{47}$V and around $I=17/2$ in
$^{49}$Cr causes the radically different rotational behavior
observed for these nuclei, namely that one signature band shows a
backbending and the other does not.

% *-------------------------------------
\section{Summary and conclusions}
\label{sect-Summary}

A comparison was made between the unpaired cranked Nilsson-Strutinsky
model and the spherical shell model.  It was found that quadrupole
properties predicted by the two models agree well. Furthermore, the
moment of inertia given by the CNS is close to that from the shell
model when either only the $T=1$ pairing is removed (even-even nuclei)
or both $T=1$ and $T=0$ pairings are removed (odd-even nuclei).

In general, the shell model gives an excellent description of
observed signature splittings. It was found that the pairing
interaction gives a strong contribution to this splitting.
Furthermore, the different spin dependence of the pairing energy
for the signature partners in \up47V and \up49Cr explains why
backbending is observed in one but not in the other signature.
This behavior comes mainly from the $T=1$ pairing but is
strengthened by the $T=0$ pairing.

Equilibrium deformations calculated in the CNS model shows that some
nuclei in the region have noticeably non-axial shapes with negative
$\gamma$-values corresponding to rotation around the intermediate
axis. These deformations can be traced back to contributions from
specific orbitals. The non-axial deformations are supported by the
calculated $B(E2)$ values and spectroscopic quadrupole moments, which
agree well with experiment as well as with those predicted by the
shell model. For \up44Ti and \up46Ti the contributions from quantum
fluctuations around the equilibrium shape lead to an improved
agreement with experimental $B(E2)$ transition strengths.

The negative-parity band in \up48Cr was discussed. It
has the largest calculated negative $\gamma$-values in this region.
It was also noted that the $B(E2)$ values
predicted by the shell model for the unstretched transitions in \up49Cr
%\cite{Brandolini01v47}
have the expected staggering behavior as a sign of triaxiality.
%\cite{Ham83}.

From this extended comparative study we conclude, that the CNS model
gives an adequate description of the quadrupole nuclear properties as
well as the occupation numbers of the spherical $j$-shells. Furthermore, 
if the pairing energy calculated from the shell model is added to the
unpaired CNS energies, excellent agreement to experimental energies is
obtained. It would be most interesting to try to include a pairing
force in the CNS model that mimics the pairing energy calculated in
the shell model. Such a model, which might be applied to all regions
of nuclei, could naturally be tested on the $pf$-shell nuclei studied
here. 

% *-------------------------------------

\begin{acknowledgments}
We would like to thank E.\ Caurier and F.\ Nowacki for access to the
shell model code {\sc Antoine} \cite{Antoine}. I.R.\ and S.\AA.\ thank
the Swedish Natural Science Research Council (NFR).
\end{acknowledgments}

% *-------------------------------------

\end{document}